# Antiferromagnetism in Pr$_3$In: Singlet/triplet physics with frustration

A. D. Christianson and J. M. Lawrence

University of California, Irvine, California 92697, USA

J. L. Zarestky

Ames Laboratory, Iowa State University, Ames, Iowa 50011, USA

H. Suzuki

National Institute for Materials Science, Tsukuba, 305-0047, Japan

J. D. Thompson, M. F. Hundley and J. L. Sarrao

Los Alamos National Laboratory, Los Alamos, New Mexico 87545, USA

C. H. Booth

Lawrence Berkeley National Laboratory, Berkeley, California 94720, USA

D. Antonio and A. L. Cornelius

University of Nevada, Las Vegas, Nevada 89154, USA

We present neutron diffraction, magnetic susceptibility and specific heat data for a single-crystal sample of the cubic (Cu$_3$Au structure) compound Pr$_3$In.  This compound is believed to have a singlet ($\Gamma_1$) groundstate and a low-lying triplet ($\Gamma_4$) excited state. In addition, nearest-neighbor antiferromagnetic interactions are frustrated in this structure. Antiferromagnetic order occurs below $T_N$ = 12K with propagation vector (0, 0, 0.5 $\delta$) where $\delta \approx 1/12$. The neutron diffraction results can be approximated with the following model: ferromagnetic sheets from each of the three Pr sites alternate in sign along the propagation direction with a twelve-unit-cell square-wave modulation. The three moments of the unit cell of 1 $\mu_B$ magnitude are aligned so as to sum to zero as expected for nearest-neighbor antiferromagnetic interactions on a triangle. The magnetic susceptibility indicates that in addition to the antiferromagnetic transition at 12K, there is a transition near 70K below which there is a small (0.005 $\mu_B$) ferromagnetic moment. There is considerable field and sample dependence to these transitions. The specific heat data show almost no anomaly at $T_N$ = 12K. This may be a consequence of the induced moment in the $\Gamma_1$ singlet, but may also be a sample-dependent effect.

61.12.Ld, 75.25.+z, 75.30.Cr, 75.40.Cx

**Introduction**

The compound $Pr_3In$ forms in the cubic $Cu_3Au$ structure (ordered fcc). Previous work[1][2][3] on polycrystalline samples of this compound indicated the existence of an antiferromagnetic transition in the range 10-20K. An additional ferromagnetic transition near 60K was observed by some authors[1][2], although others[3] argued that this transition was due to a secondary $Pr_2In$ crystal phase. There are two interesting aspects of the physics of this compound. First, although the Pr site symmetry is tetragonal, the crystalline electric field at the Pr site is believed to have nearly-cubic symmetry[3][4] and a

$\Gamma_1$ singlet ground state and a low-lying $\Gamma_4$ triplet excited state. The isostructural compound Pr$_3$Tl, for which similar statements hold, was studied extensively as a classic singlet/triplet induced moment ferromagnet[4,5,6,7]. In such systems, when the intersite exchange interaction is sufficiently large relative to the singlet/triplet splitting, a phase transition occurs such that for T < T$_C$ a moment is induced in the ground state singlet via admixture with the triplet states. Such induced order is also expected in Pr$_3$In. Secondly, in this structure, where the Pr atoms form triangular lattices perpendicular to the [1,1,1] direction, nearest neighbor antiferromagnetic (AF) interactions should be frustrated[8]. Hence, this compound may be a frustrated, induced moment singlet/triplet antiferromagnet.

In this paper, we report neutron diffraction, magnetic susceptibility and specific heat results for a single-crystal of Pr$_3$In. We give a preliminary model of the antiferromagnetic structure that aligns the moments in the unit cell as expected for frustrated AF interactions. We use the magnetic susceptibility to show that there is also a weak ferromagnetic component of the magnetic structure, as well as considerable sample and field dependence to the results. We use the specific heat data to strengthen the case for singlet/triplet induced magnetism.

**Experimental details**

A large (1 cm diameter by 2 cm long) cylindrical boule of Pr$_3$In was grown by the Bridgman technique using a Mo crucible. The neutron diffraction measurements were performed on the Ames Laboratory triple-axis spectrometer, HB1A, at the High Flux

Isotope Reactor (HFIR) at the Oak Ridge National Laboratory (ORNL). The HB1A spectrometer operates with a fixed initial energy of 14.7 meV using a double pyrolitic graphite monochromator system. This, together with two HOPG filters, provides a very intense and clean neutron beam ($I_{\lambda/2} \sim 10^{-4} I_\lambda$). A pyrolitic graphite analyzer and collimations of 48'-40'-40'-102' were also used. The sample contained a large (~ 1 cm$^3$) irregularly-shaped crystal that was not aligned with the growth axis of the boule, and several smaller crystallites. We performed measurements for both the (hhl) and (hk0) reciprocal lattice planes of the large crystal oriented in the scattering plane of the spectrometer. Because of the irregular shape and orientation of the crystal, we were unable to correct for sample absorption, which was significant due to the large absorption cross section of In. The units of intensity given below are counts per monitor count units (1 mcu ° 1 sec). The susceptibility and specific heat measurements were performed on two small pieces cut from the center of the large crystallite; results of these measurements on the two pieces were identical. The susceptibility was measured in commercial (Quantum Design) SQUID magnetometers at Los Alamos National Laboratory (LANL) and Lawrence Berkeley National Laboratory (LBNL); the specific heat was measured via a thermal relaxation method using commercial (Quantum Design PPMS) systems at LANL and Las Vegas.

## Experimental results and analysis

The low temperature neutron diffraction results confirmed the Cu$_3$Au crystal structure with lattice constant 4.94Å. The inset to Fig. 1c displays the experimental structure factor $F^2_{exp} = I \sin\theta$, i.e. the intensities multiplied by the Lorentz factor $sin\theta$ appropriate for $q$-scans[9], for several nuclear peaks in the (1, -1, 0) scattering plane. The upper solid line is the average value ($F^2_{av}$ = 13510) of the experimental structure factor for the high intensity peaks; the lower solid line is the expected value (13510 $[(b_{Pr} - b_{In})/(3b_{Pr} + b_{In})]^2$ = 11) of the structure factor for the low-intensity peaks. The fact that the measured intensities of the low intensity peaks cluster around this latter value suggests that the Cu$_3$Au crystal structure is well-ordered. The deviations from the average values arise primarily from (uncalculable) absorption effects, but may also reflect the inadequacy of the $sin\theta$ approximation to the Lorentz factor[9]. Extinction effects, expected for such a large sample, would imply that the high-intensity peaks should be intrinsically stronger than measured, so that the low-intensity peaks would be relatively weaker. Given these observations, it is not possible to rule out some degree of site disorder or variation of the stoichiometry from the 3:1 ratio.

The susceptibility, measured in a magnetic field of 0.01T, is shown in Fig. 2. The peak at 12K indicates the onset of antiferromagnetic order. The high temperature susceptibility (Fig. 2a, inset) can be approximated by a Curie-Weiss law $c = C(Pr)/(T - q)$ where $C(Pr)$ = 1.55emu/mol-Pr is the free-ion Curie constant for Pr ($J = 4$); the approximation is particularly good in the range 100-200K. The value $q$ = 12K suggests ferromagnetic interactions. A small jump occurs in the susceptibility at 70K. This jump is seen more clearly in a plot of the effective moment $T c/C(Pr)$, which approaches the free-ion value of unity at high temperature, but which increases dramatically below 70K (Fig. 2c). This increase is a clear sign of ferromagnetism; the decrease at low temperatures arises both from saturation of the ferromagnetic contribution and from the onset of antiferromagnetism. Plots of the magnetization (Fig. 3) show hysteresis below 70K. Both the coercive field (0.015T) and the remanent magnetization (0.005$m_B$) are very small at $T = 5K$. Hence, the ferromagnetism which occurs below 70K is very weak in this compound.

Fig. 4 shows that the effect of increasing the magnetic field is to decrease the temperature of the susceptibility maximum (Fig. 4a and inset). In the effective moment plots (Fig. 4b and inset) it can be seen that increasing the magnetic field decreases the magnitude of the discontinuity at 70K. The susceptibility for a piece cut from the end of the Bridgman boule (and thus outside the region of the single crystallite used in the neutron measurement) is shown in both panels; the overall magnitude is similar to that of the centerpiece, but there is no sign of antiferromagnetic order. This probably reflects a difference in stoichiometry of the endpiece relative to that of the large crystallite.

The specific heat data is shown in Fig. 5. The lattice contribution was determined from previous measurements[10] of La$_3$In; the temperature-dependent Debye temperature $Q_D(T)$ given in that paper was extrapolated in linear fashion to higher temperature (T > 16K) and then used to evaluate the Debye specific heat. The magnetic specific heat then was taken as the measured value minus the lattice contribution. The upturn in the data at the lowest temperatures is from a contribution of the Pr nucleus due to a large hyperfine field, in agreement with Ref. 3. The magnetic specific heat and the corresponding entropy is very small at 12K and the specific heat anomaly associated with the antiferromagnetic transition is so weak as to be only barely visible in a plot of $C_{mag}/T$

(Fig. 5b, inset). No sign of an anomaly in the specific heat was observed near 70K, where the susceptibility exhibits a discontinuity.

Fig. 1 shows the magnetic reflections (marked M) observed below 12K in the neutron diffraction for the (1, -1, 0) scattering plane. Strong (e.g. (0, 0, 2)) and weak (e.g. (1, 1, 2)) nuclear reflections (marked N) are also present, as well as peaks (marked Al) arising from polycrystalline aluminum in the sample environment. The magnetic peaks can be indexed as occurring at $q = (h, k, l + ½ ± d)$ where $d = 0.083 ≈ 1/12$. (We cannot rule out that the ordering is slightly incommensurate.) Similar results were seen in the (0, 0, 1) scattering plane, where peaks were observed at $(h + ½ ± d, k, 0)$ and $(h, k + ½ ± d, 0)$. In addition to these primary reflections, several 3d harmonics were observed, e.g. the peak marked H at (2, 2, 1.75) seen in Fig. 1c. Finally, a single 4d harmonic was observed at (1, 1.16, 0). The primary magnetic peaks and the 3d and 4d harmonics vanish above 12K and the temperature dependence of the 1d, 3d and 4d reflections can be approximated as $B_i + C_i [(T_N - T)/T_N]^{1/2}$ where $T_N = 11.4K$ but where the backgrounds $B_i$ and coefficients $C_i$ are different for the different reflections (Fig. 6). Hence the harmonics have the same temperature dependence as the order parameter, which varies in a manner typical of an antiferromagnetic transition. In addition, there may be a small variation in the value of the ordering wavevector, i.e. $d$, between 10 and 12K (Fig. 6b, inset), suggesting that the wavevector is initially incommensurate, but then locks on to the commensurate value 1/12 below 10K.

The magnetic line intensities observed in Fig. 1 can be approximated by the following model (Fig. 7, insets). Each of the three Pr sites in the unit cell ($r_1$ = (½, ½, 0), $r_2$ = (0, ½, ½) and $r_3$ = (½, 1, ½)) gives rise to a sublattice ($r_i$ + (h, k, l)) of spins consisting of ferromagnetic sheets perpendicular to the propagation (z-) direction which alternate in sign along the z-direction. In and of itself this would yield a basic two-unit cell structure with $q$ = (0, 0, ½). The direction of the moments is, however, modulated by a twelve unit cell square wave so that the magnetic reflections occur at $z = ½ \pm (2n+1)d$, where $d = 1/12$. The direction of the moments in the first unit cell is taken as $\hat{S}_1$ = (1, 0, 0), $\hat{S}_2$ = (-1/2, 0, -√3 /2), and $\hat{S}_3$ = (-1/2, 0, +√3 /2) and the magnitude is $1m_B$. The results of a calculation of the diffraction intensities for this structure are compared to the measured magnetic reflection intensities in Fig.7. The calculated intensities have been normalized to the experimental nuclear structure factor (Fig. 1c, inset) to facilitate direct comparison to the experimental data. The intensities are modulated as a function of angle by the Pr form factor and the Lorentz factor, taken again as $sinq$.

Fig. 7 shows that this model gives a good first approximation to the line intensities. It reproduces the alternation of intensities along and between (h, h, l) lines and it approximates the magnitudes fairly well. Some of the predicted harmonics (e.g. (2, 2, 1.75)) are observed at about the correct intensity. To an unknown extent, the discrepancies between the measured and predicted line intensities can be attributed to the same sources as the deviations seen in Fig. 1c, inset, especially the uncalculable absorption correction.

For several reasons, we view this model only as a reasonable starting point in describing the antiferromagnetic structure. First, it is clear from the susceptibility that there is a small ferromagnetic component in the structure. (Given the very small magnitude of this component, $0.005 m_B$ as deduced from Fig. 3, its effects on the line intensities could not be resolved, given the statistics of this experiment.) Second, in this structure no even harmonics are expected, whereas a small-intensity 4d harmonic was observed at (1, 1.16, 0) with a temperature dependence (Fig. 6) proportional to the order parameter. Finally we note that other alignments of spins (e.g. $\hat{S}_1 = (1, 0, 0)$, $\hat{S}_2 = (0, 1, 0)$ and $\hat{S}_3 = (0, 0, 1)$), such that the sum $(S_2 + S_3)_{xy}$ of the projections of the two moments at $r_2$ and $r_3$ onto the $xy$-plane is $1 m_B$, give essentially similar results to those of Fig. 7. (However, variation of $(S_2 + S_3)_{xy}$ away from the value $1 m_B$ significantly degrades the comparison to the experiment.)

## Discussion

We first consider crystal quality and the sample dependence of these results. Based on the nuclear line intensities, where the weak lines that are forbidden in the pure fcc structure have roughly the right intensity relative to the strong lines (Fig. 1c, inset) it is clear that our crystal is reasonably well-ordered in the $Cu_3Au$ structure. As mentioned, however, given the uncertainties due to absorption and extinction, we cannot rule out some degree of disorder or deviation from the correct 3:1 stoichiometry. Past studies of $Pr_3In$[1,2,3] show considerable variation in the magnitudes of the antiferromagnetic and the ferromagnetic contributions to the susceptibility, implying that sample quality is an important issue in this compound. Given that Pr$_3$In is slightly peritectic[11] and does not grow congruently from the melt, it is reasonable to assume that samples grown from the melt, either as arc-melted polycrystals or as Bridgman-grown single crystals, will deviate somewhat from the correct 3:1 stoichiometry. This is probably the main source of disagreement between results on different samples. We note that the susceptibility of a piece cut from the end of our sample (Fig. 4) shows no antiferromagnetic transition, which probably results from a stoichiometry variation between the outer edges and the center of the boule, where the large single crystal was located. We note also that the ferromagnetic anomaly in the susceptibility at 70K in our samples is considerably smaller than that seen in other studies, with the exception of Ref. 3, where no such anomaly was reported. However, the field used in the latter study (1.5T) was sufficiently large that (given the field dependence shown here in Fig. 4) the anomaly may have been suppressed. Since the samples that we used for the susceptibility measurement were cut from the center of the single crystal, we believe that the ferromagnetic anomaly is intrinsic to Pr$_3$In, and not due to the presence of a second Pr$_2$In phase, as suggested by Ref. 3. Given the uncertainties in stoichiometry and Cu$_3$Au site order, the intrinsic strengths of the feromagnetic and antiferromagnetic contributions remain uncertain.

The very small anomaly in the specific heat at the antiferromagnetic transition is quite striking, especially given the well-defined temperature dependence of the order parameter, which is typical for an antiferromagnetic transition. In an early study[4] of the classic singlet/triplet ferromagnet Pr$_3$Tl, a very small specific heat anomaly was also observed. This was attributed to the fact that in a mean-field treatment of the induced-moment ferromagnet, there is no change in entropy in the $\Gamma_1$ singlet at the transition, but rather the singlet, which has no moment above $T_C$, acquires a moment from admixture with the $\Gamma_4$ states below $T_C$. In a more recent study[5] of $Pr_3Tl$, however, a well-defined specific heat anomaly was observed. This was attributed to entropy arising from low-lying magnetic modes that go soft at the transition and which are not included in the mean-field theory. Unfortunately, these low-lying modes have never been observed experimentally[6,7]. In any case, it is clear that the specific heat in $Pr_3Tl$ is sample dependent, so that sample dependence of the specific heat of $Pr_3In$ should also be expected. Hence we cannot be certain that the lack of an anomaly in the specific heat is intrinsic to $Pr_3In$, but it does seem to be a common feature of systems exhibiting induced moment magnetism.

Turning now to the magnetic structure, we note that the core of the model proposed above is that the three spins in the unit cell point along the edges of an equilateral triangle, and therefore sum to zero. This is the lowest energy state for the simpler case of three antiferromagnetically-coupled spins on an equilateral triangle[8]. In the present case, ferromagnetic next-nearest-neighbor (nnn) interactions stabilize ferromagnetic sublattices of these three nearest neighbor (nn) spins. Indeed, were there only nn antiferromagnetic and ferromagnetic nnn interactions, the lowest state would be a $q = 0$ structure with all unit cells identical to the core cell. The complicated structure that we observe, with the sign of the moments alternating between neighbor cells along the propagation direction, and further modulated by the 12-unit-cell square wave, must arise from longer range interactions.

As discussed above, models where the two spins in the base unit cell at $z = ½$ have total projection onto the (0, 0, 1) plane equal to that of the (½, ½, 0) spin give equally good fits to the magnetic reflection intensities. Indeed, three antiferromagnetically-coupled moments placed on a triangular lattice is the paradigm case of frustration, and many other low-lying states are possible. Different patterns of order could then be stabilized in the presence of competing interactions (nnn, nnnn, etc.). In addition, an extension of the model is required to explain the existence of the even order harmonic at (1, 1.16, 0). Furthermore, the model needs to account for the weak ferromagnetism that sets in below 70K. The small saturation moment (Fig. 3) probably reflects a small canting of the moments, giving a *ferri*magnetic component to the order. It is also probable that the small magnitude of this component reflects the singlet/triplet physics of this compound. Given all this, we take our model structure as a first approximation to the antiferromagnetic order in $Pr_3In$.

Finally, we note that, in a purely frustrated system with no competing, stabilizing interactions, entropy generation would be spread over a large temperature range as low-lying modes of order were excited. Perhaps a remnant of this effect is partly responsible for the vanishingly small entropy change at $T_N$. On the other hand, frustrated antiferromagnets usually exhibit a large value of the ratio $Q_{CW}/T_N$ where $Q_{CW}$ is the antiferromagnetic Curie-Weiss parameter, whereas in the present case $Q_{CW}$ is ferromagnetic and essentially equal to $T_N$. While this suggests that the effects of frustration may be negligible here, it is not obvious to us that this criterion is valid when the frustrated antiferromagnetism is *induced* in a ground state singlet.

## Conclusion

We have shown that antiferromagnetic order with primary reflections at $q = (h, k, l + ½ \ 0.083)$ occurs below 12K in $Pr_3In$. There appears to be very little entropy change associated with the transition. In addition, a weak ferromagnetic component sets in near 70K. There is considerable sample dependence to these effects. The physics appears to combine singlet/triplet induced moment magnetism and frustrated nearest-neighbor antiferromagnetic interactions. Longer range interactions, both ferromagnetic and antiferromagnetic, are clearly significant. Further experiments are needed to determine the sample dependence and the intrinsic behavior, and to refine the magnetic structure. The crystal-field level structure needs to be determined directly by neutron scattering to prove that the singlet/triplet $\Gamma_1/\Gamma_4$ model is applicable. Finally, given that the soft dispersive crystal field modes expected near the transition in the singlet/triplet model have not been observed experimentally, even in the simpler ferromagnetic case of $Pr_3Tl$,[6] [7]measurement of the spin dynamics is a crucial future experiment.

## Acknowledgments

We thank Steve Shapiro, Eric Bauer and Jason Gardner for helpful discussions. Work at UC Irvine was supported by the U. S. Department of Energy (DOE) under Grant No. DE-

FG03-03ER46036. Work at UNLV was supported by DOE EPSCoR-State/National Laboratory Partnership Award DE-FG02-00ER45835 and Cooperative Agreement DE-FC08-01NV14049. Work at LBNL was supported by the Director, Office of Science, Office of Basic Energy Sciences, of the U.S. DOE under Contract No. DE-AC03-76SF00098. Ames Laboratory is operated by Iowa State University for the U.S. Department of Energy (DOE) under Contract No. W-7405-ENG-82. Work performed at the HFIR Center for Neutron Scattering was supported by the DOE Office of Basic Energy Sciences Materials Science, under Contract No. DE-AC05-000R22725 with UT-Battelle, LLC. Work at Los Alamos was performed under the auspices of the DOE.

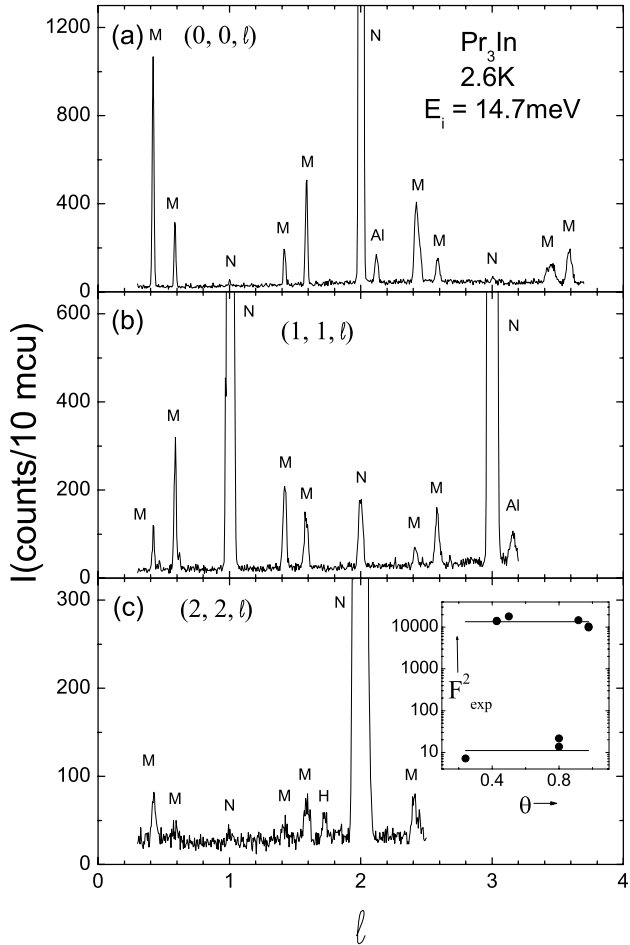

Fig. 1 Neutron diffraction profiles for Pr$_3$In at 2.6K along three lines in the (1, -1, 0) scattering plane. Peaks marked M are the primary magnetic reflections that appear below 12K. Peaks marked N are nuclear reflections and peaks marked Al are due to polycrystalline aluminum in the beam. The peak marked H is a harmonic of the primary magnetic reflections. Inset: The experimental structure factor *Isinq* for the nuclear peaks in the (1, -1, 0) scattering plane. The upper solid line is the average value for the strong peaks; the lower solid line is the predicted value for the weak peaks based on the average value for the strong peaks.

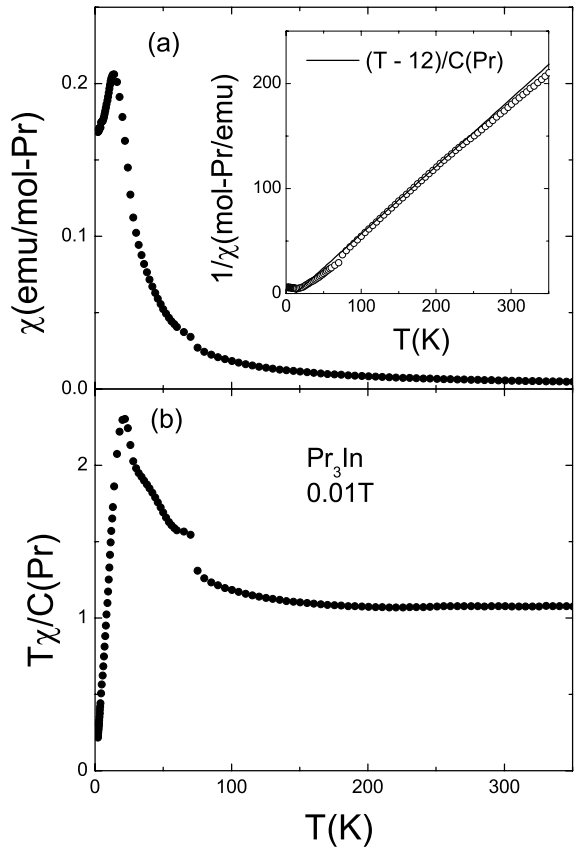

Fig. 2  a) The susceptibility $\chi(T)$ of $Pr_3In$ measured with H = 0.01T for a piece cut from the center of the single crystal.  The inset compares the inverse susceptibility to a Curie-Weiss law.  b) The effective moment $T\chi/C(Pr)$ where $C(Pr)$ is the Pr free-ion Curie constant.

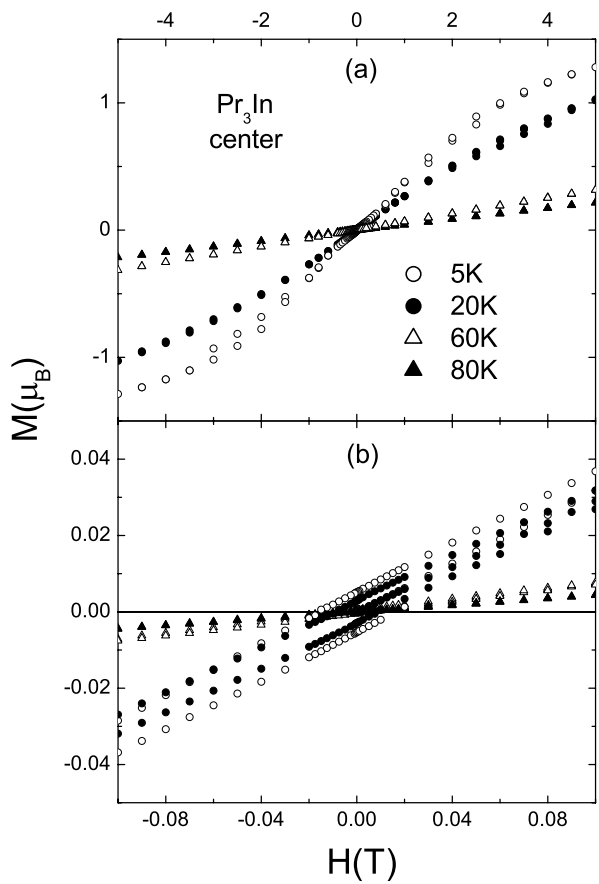

Fig. 3 a) The magnetization of Pr$_3$In at four temperatures above and below the transition at 70K. Data taken with both field increasing and field decreasing are included to establish the hysteresis that occurs below 70K. b) Magnetization shown on an expanded scale.

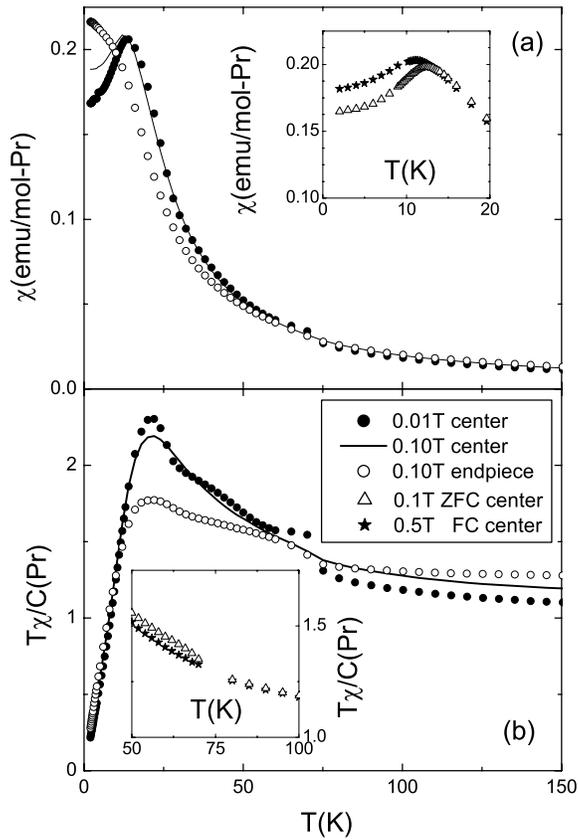

Fig. 4 a) The susceptibility at *H* = 0.01 and 0.1T for a piece cut from the center of the single crystal of Pr₃In together with susceptibility for a piece cut from the end of the Bridgman boule. The inset emphasizes the effect of increasing magnetic field on the susceptibility near the antiferromagnetic transition. b) The effective moment *T*χ*/C(Pr)* under the same conditions as in a); the inset emphasizes the effect of increasing magnetic field on the susceptibility near the transition at 70K.

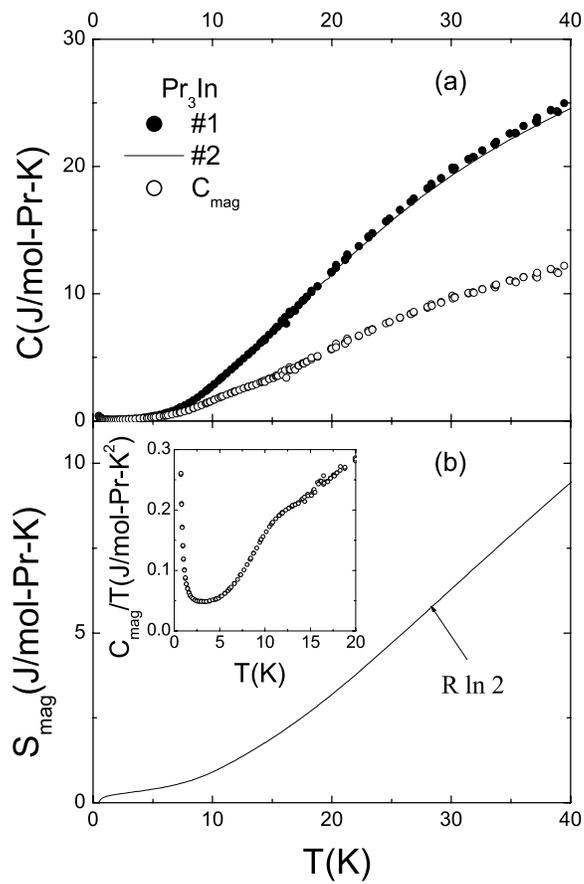

Fig. 5 a) The specific heat for two samples cut from the center of the single crystal of Pr$_3$In. The open circles exhibit the magnetic contribution, determined as discussed in the text. b) The entropy associated with the magnetic specific heat. The inset shows the linear coefficient $C_{mag}(T)/T$.

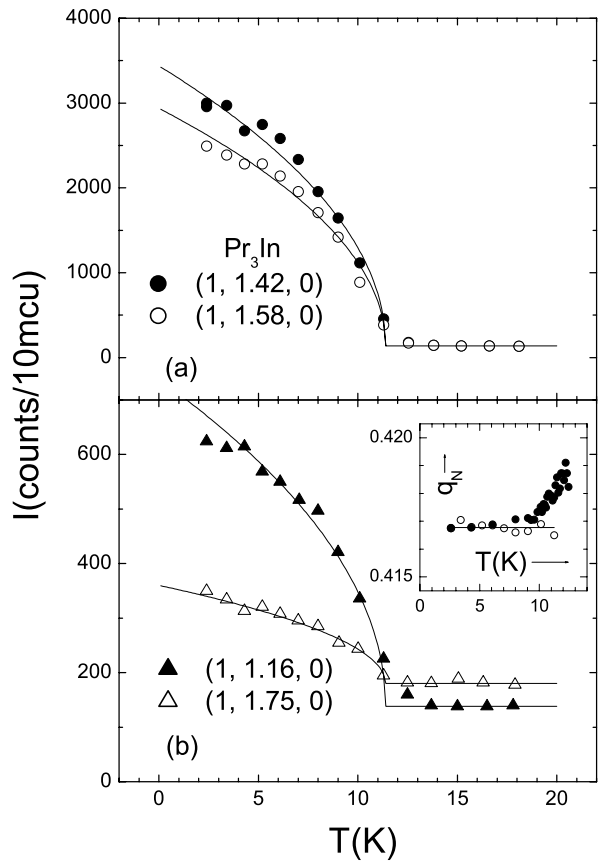

Fig. 6 The intensity in the primary magnetic reflections (1, 1.42, 0) and (1, 1.58, 0) (a) and in the 4d (1, 1.16,0) and 3d (1, 1.75, 0) harmonics (b) vs. temperature. The solid lines represent the behavior $B_i + C_i \, [(T_N - T)/T_N]^{1/2}$ with $T_N = 11.4$K and with $B_i$ and $C_i$ varying between reflections.

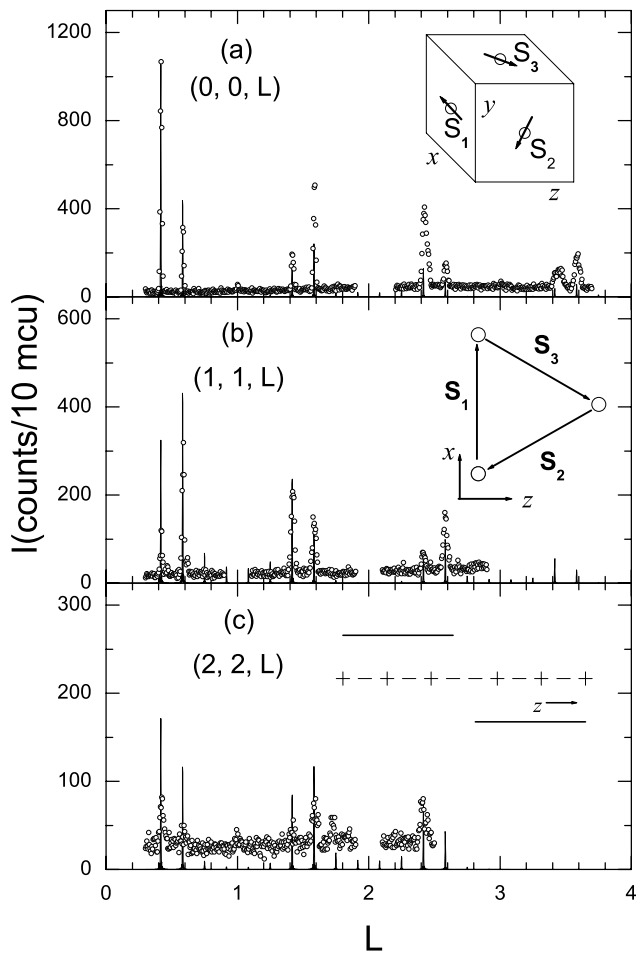

Fig. 7  The measured magnetic reflections (open circles) compared to the intensities calculated for the model of the antiferromagnetic order given in the text (solid lines).  The insets depict the model:  a) Three Pr moments $S_i$ of magnitude $1m_B$ sit on the face centers of the unit cell; b) the moments lie in the $xz$ plane at angle 120± with respect to each other and sum to zero as shown; c) each $S_i$ forms a sublattice that consists of ferromagnetic sheets in the $xy$ plane that alternate in sign along the $z$-direction within the envelope of a 12-unit cell square wave.